\documentclass[pra,aps,floatfix,twocolumn,superscriptaddress,amsmath,amssymb]{revtex4-2}
\usepackage{graphicx}
\usepackage{color}
\usepackage{bm}
\usepackage{hyperref}
\hypersetup{colorlinks=true,citecolor=blue,linkcolor=blue,urlcolor=blue}
\usepackage[T1]{fontenc}
\usepackage{newtxtext,newtxmath}
\usepackage[normalem]{ulem}
\allowdisplaybreaks

\begin{document}

\title{%
Ground-state phase diagram of the SU($4$) Heisenberg model on a plaquette lattice
}

\author{Ryui Kaneko}
\email{ryuikaneko@sophia.ac.jp}
\affiliation{%
Physics Division, Sophia University, Chiyoda, Tokyo 102-8554, Japan}

\author{Shimpei Goto}
\email{goto.las@tmd.ac.jp}
\affiliation{College of Liberal Arts and Sciences, Tokyo Medical and Dental University, Ichikawa, Chiba 272-0827, Japan}

\author{Ippei Danshita}
\email{danshita@phys.kindai.ac.jp}
\affiliation{%
Department of Physics, Kindai University, Higashi-Osaka, Osaka 577-8502, Japan}

\date{\today}

\begin{abstract}
We investigate the ground state of the SU($4$) Heisenberg model on a
square lattice with spatial anisotropy on each plaquette bond
using the tensor-network method based on infinite projected entangled
pair states.
We find that the SU($4$) singlet ground state appears in the strongly
anisotropic limit, whereas 
N\'eel and valence-bond crystal orders coexist
in the nearly isotropic limit. By examining the intermediate
parameter region, we identify a phase transition between these phases.
The nature of the phase transition is likely to be of first order,
and the transition point is estimated to be around
$J'/J\approx 0.85(5)$, where $J$ and $J'$ are the interaction strengths of
intra- and interplaquette bonds, respectively.
We also calculate the anisotropy dependence of singlet correlations on a plaquette bond,
which will be useful for future experiments
of ultracold atoms in optical lattices.
\end{abstract}

\maketitle

\section{Introduction}
\label{sec:introduction}

Lattice models with SU($N$) symmetry, where $N > 2$, have recently attracted
significant interest owing to the potential emergence of novel quantum
states not found in the Hubbard and Heisenberg models with conventional
SU($2$) symmetry~\cite{wu2003,honerkamp2004,cazalilla2009,gorshkov2010,
hermele2011,yamamoto2024}.
About a dozen years ago,
such SU($N>2$) systems were experimentally realized
with ultracold atoms in optical lattices
using $^{173}$Yb~\cite{taie2012}, which is a fermionic isotope of alkaline-earth-like atoms.
In contrast to other realistic platforms of SU($N>2$) systems,
such as antiferromagnets with coupled spin and orbital degrees of
freedom~\cite{kitaoka1998,reynaud2001,nakatsuji2012,
quilliam2012,smerald2014,
kugel2015,yamada2018},
ultracold atoms in optical lattices are highly controllable.

Observing antiferromagnetic correlations in
optical-lattice systems has
long been a central issue in ultracold-atom
experiments~\cite{greif2013,hart2015,mazurenko2017,ozawa2018,shao2024_arxiv}
and recent advances in
cooling techniques have enabled the observation of
antiferromagnetic correlations in SU($N$) systems
for large $N$~\cite{taie2022}.
Remarkably,
antiferromagnetic correlations are more enhanced
for SU($6$) systems than for SU($2$) systems,
thanks to strong Pomeranchuk cooling effects~\cite{taie2022}.
This finding further stimulates research on
quantum phase transitions caused by antiferromagnetic order
in SU($N$) systems with much larger $N$.
On the other hand, antiferromagnetic correlations in SU($N$) systems
for rather small $N=2$ and $4$
can also be experimentally investigated
on a cubic lattice
with a spatial anisotropy in the hopping amplitudes
favoring dimerization~\cite{ozawa2018}.

The physics of SU($N>2$) systems is drastically different from
that of SU($2$) systems, even for a relatively small $N$,
because of intertwined spin and orbital degrees of freedom.
For example, the ground states of the SU($3$) Heisenberg models
exhibit a variety of exotic phases
depending on the lattice geometry and the strength of the interactions.
Spin nematic order emerges on a triangular lattice~\cite{tsunetsugu2006,
lauchli2006,yamamoto2020},
unconventional stripe order appears on a square lattice~\cite{toth2010},
and plaquette valence-bond crystal (VBC) orders
or related competing orders are found on honeycomb~\cite{lee2012,
zhao2012,corboz2013}
and kagome~\cite{arovas2008,corboz2012a}
lattices.

\begin{figure}[t]
\centering
\includegraphics[width=0.5\columnwidth]{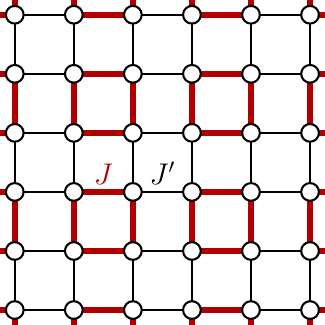}
\caption{%
Square lattice with spatial anisotropy on plaquette bonds.
The interaction strength of intra- and interplaquette bonds is denoted by
$J$ (thick red lines) and $J'$ (thin black lines), respectively.
When $J'/J=0$ and $J/J'=0$,
the system decouples into isolated four-site chains.
}
\label{fig:lattice}
\end{figure}

However, the numerical investigation of quantum many-body states
in SU($N$) systems becomes significantly challenging when $N$
increases~\cite{miyazaki2021,miyazaki2022,corboz2011b,
li1998,vandenbossche2000,vandenbossche2001,hung2011,szirmai2011,
corboz2012b,lajko2013,mikkelsen2023,yu2024}.
Even for the SU($4$) Heisenberg model on a simple lattice,
the ground state is still under debate.
For instance,
cluster mean-field and spin-wave
approximations have been used to investigate the ground state of the
SU($4$) Heisenberg model on a plaquette lattice, i.e., a square lattice with
spatial anisotropy on plaquette bonds
(see Fig.~\ref{fig:lattice})~\cite{miyazaki2021,miyazaki2022}.
These studies found that the SU($4$)
singlet ground state is favored in the strongly anisotropic regime. In the
opposite limit, a prior tensor-network study suggested that
the ground state should exhibit coexisting N\'eel and
VBC order in the nearly isotropic region~\cite{corboz2011b},
although there would be many competing ground-state candidate
states~\cite{li1998,vandenbossche2000,vandenbossche2001,hung2011,szirmai2011}.
These findings imply a phase
transition or other phases in the intermediate anisotropy region.

To understand the interplay between lattice spatial anisotropy and the
spontaneous breaking of spin rotational and lattice translational
symmetries in a simple SU($N$) system,
we investigate the ground states of the SU($4$) Heisenberg
model on a plaquette lattice using a two-dimensional tensor-network
method based on infinite projected entangled pair states
(iPEPS)~\cite{martin-delgado2001,verstraete2004_arxiv,verstraete2004,
verstraete2008,jordan2008,phien2015,orus2014,orus2019}
or infinite tensor product states~\cite{hieida1999,okunishi2000,nishino2001,
maeshima2001,nishio2004_arxiv}.
We successfully reproduce the SU($4$) singlet ground state in the strongly
anisotropic limit and the N\'eel-VBC coexisting ground state in the nearly
isotropic limit. By examining the intermediate parameter region, we
identify a phase transition between these phases, which will be relevant
to future experimental investigations.

This paper is organized as follows:
In Sec.~\ref{sec:model}, we introduce the SU($4$) Heisenberg model on a
plaquette lattice and the tensor-network method used in this study.
In Sec.~\ref{sec:results}, we present the results of the ground-state
calculations and discuss the phase diagram of the model.
We also calculate the anisotropy dependence of singlet correlations on
plaquette bonds
that can be measured in future experiments with ultracold atoms in
optical lattices.
Finally, we summarize our findings and discuss future prospects in
Sec.~\ref{sec:summary}.
We set the reduced Planck constant as $\hbar=1$ and
a lattice constant as $a=1$
throughout this paper.

\section{Model and method}
\label{sec:model}

We consider the SU($4$) antiferromagnetic Heisenberg
model~\cite{corboz2011b},
\begin{align}
 H_0
 = J  \sum_{\langle ij \rangle_{\rm intra}} \hat{P}_{ij}
 + J' \sum_{\langle ij \rangle_{\rm inter}} \hat{P}_{ij}
\end{align}
on a plaquette lattice~\cite{miyazaki2021}
(see Fig.~\ref{fig:lattice}).
Here, $\hat{P}_{ij}$ is a transposition operator
which exchanges flavors at sites $i$ and $j$, namely,
$\hat{P}_{ij}|\alpha_i \beta_j\rangle = |\beta_i \alpha_j\rangle$
($\alpha_i, \beta_i \in \{0,1,2,3\}$).
The symbols $\langle ij \rangle_{\rm intra}$
and $\langle ij \rangle_{\rm inter}$
denote nearest-neighbor sites
within a plaquette and between neighboring plaquettes, respectively.
The interaction strengths for intra- and interbonds
are denoted by $J$ and $J'$, respectively.
Using the flavor-changing operator
$\hat{S}^{\beta}_{\alpha}(i)=|\alpha_i\rangle\langle\beta_i|$,
the Hamiltonian can be written as
\begin{align}
 H_0
 =
   J  \sum_{\langle ij \rangle_{\rm intra}}
   \sum_{\alpha\beta} \hat{S}^{\beta}_{\alpha}(i) \hat{S}^{\alpha}_{\beta}(j)
 + J' \sum_{\langle ij \rangle_{\rm inter}}
   \sum_{\alpha\beta} \hat{S}^{\beta}_{\alpha}(i) \hat{S}^{\alpha}_{\beta}(j).
\end{align}
Since the model with $J>J'$ and that with $J<J'$
are equivalent by interchanging the interaction strengths,
we investigate the ground state of the model
by controlling $J'/J \in [0,1]$.

At $J'/J=0$,
the system decouples into four-site chains,
and the ground state is an SU($4$) singlet on each plaquette
bond~\cite{miyazaki2021,miyazaki2022}.
On the other hand, at $J'/J=1$,
the ground state is likely to be a N\'eel-VBC coexisting
state~\cite{corboz2011b}.
Because both states are expected to have nonzero excitation gaps
(of tetramer and dimer orders),
these states would be robust against small perturbations.
Then, the ground state would be a tetramerized state for $0<J'/J\ll 1$, 
where as it would still be the VBC state
for $0<1-J'/J\ll 1$.
However, to the best of our knowledge,
the nature of the ground state in the intermediate region ($0<J'/J<1$)
has not been clarified yet.

\begin{figure}[t]
\centering
\includegraphics[width=0.9\columnwidth]{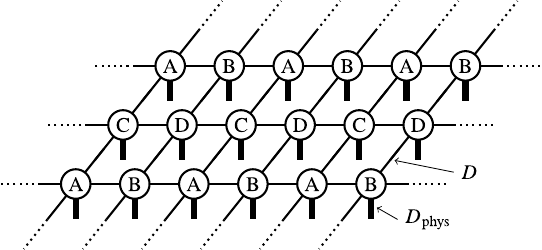}
\caption{%
Schematic representation of the iPEPS structure.
Here we take a $2\times 2$ sublattice structure
as an example.
The corresponding tensors $A$ and $B$ are
rank-five tensors denoted by the circles.
The tensors are connected
with the physical (thick lines) and virtual (thin lines) bonds.
The bond dimensions for the physical and virtual bond
are denoted by $D_{\rm phys}(=4)$ and $D$, respectively.
}
\label{fig:ipeps}
\end{figure}

One needs to take into account the effect of quantum fluctuations
more accurately
to investigate the stability of such crystal states
beyond the mean-field-level approximations.
To this end,
we apply the two-dimensional tensor-network method
based on iPEPS~\cite{martin-delgado2001,verstraete2004_arxiv,verstraete2004,
verstraete2008,jordan2008,phien2015,orus2014,orus2019,
hieida1999,okunishi2000,nishino2001,
maeshima2001,nishio2004_arxiv}.
We illustrate the schematic structure
of iPEPS in Fig.~\ref{fig:ipeps}.
The physical bond dimension is chosen as $D_{\rm phys}=4$,
corresponding to the four flavors $\alpha=0,1,2,3$.

The ground-state candidate state for $J'/J\ll 1$, namely,
the SU($4$) singlet state,
can be represented by the iPEPS with the bond dimension
$D=7$.
On the other hand, for $J'/J\approx 1$,
the ground state is expected to be a N\'eel-VBC coexisting state,
which is found to be
prepared
after optimizing the dimerized initial state
in iPEPS calculations
for the bond dimension $D\ge 3$.
For the details of the initial state preparation,
see the Appendix~\ref{sec:initial_state}.

\begin{figure}[t]
\centering
\includegraphics[width=0.45\columnwidth]{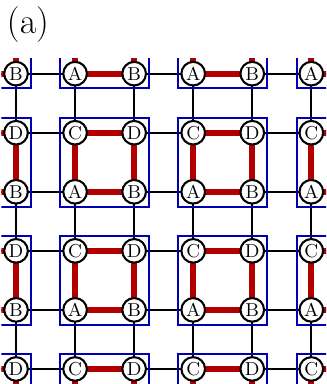}
\hfil
\includegraphics[width=0.45\columnwidth]{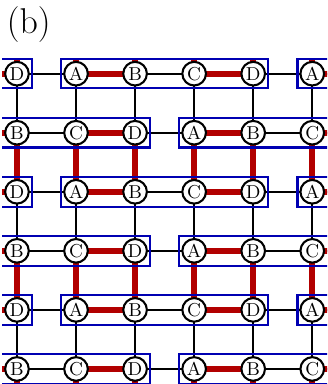}
\bigskip
\\
\includegraphics[width=0.45\columnwidth]{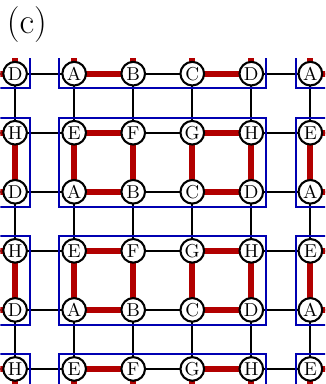}
\hfil
\includegraphics[width=0.45\columnwidth]{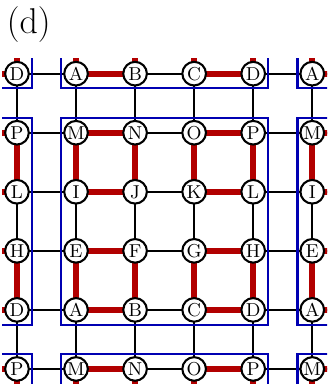}
\caption{%
Sublattice structures for the iPEPS calculation.
We choose
(a) $2\times 2$ sublattice,
(b) $4\times 1$ sublattice,
(c) $4\times 2$ sublattice, and
(d) $4\times 4$ sublattice
structures.
}
\label{fig:sublattice}
\end{figure}

We take the aforementioned SU($4$) singlet and
dimerized states as initial states
and optimize these wave functions
under several sublattice structures
via imaginary-time evolution,
as shown in Fig.~\ref{fig:sublattice}.
The $2\times 2$ sublattice structure
is suitable for representing the SU($4$) singlet state,
while the $4\times 1$ sublattice structure
is required for the N\'eel-VBC coexisting state.
The $4\times 2$ sublattice structure
includes both $2\times 2$ and $4\times 1$ sublattice structures
and allows for representing both ground-state candidate
states and their coexistence, if any.
To carefully take into account the effect of quantum fluctuations
and investigate the possibility of other complicated ordered states,
we also consider the $4\times 4$ sublattice structure,
which includes the $4\times 2$ sublattice structure.
Moreover, we use several random initial states
when the SU($4$) singlet and N\'eel-VBC coexisting states
are competing for $D=7$.
After the optimization, we have always found that
the random initial state converges to the N\'eel-VBC coexisting state.
Therefore, we mainly optimize
the SU($4$) singlet and N\'eel-VBC coexisting states
for bond dimensions $D>7$.

We adopt the TeNeS library~\cite{motoyama2022,tenes,ptns}
and calculate the ground-state candidates up to $D=12$
using the simple update algorithm~\cite{jiang2008,jordan2008}.
We calculate physical quantities in the thermodynamic limit
using the corner transfer matrix renormalization group (CTMRG)
method~\cite{nishino1996,nishino1997,
nishino1999,okunishi2000,
orus2009,corboz2010,
corboz2011a,corboz2014,phien2015,orus2014,orus2019}.
To ensure the convergence of the physical quantities,
we choose the bond dimension of the environment tensors
as $\chi=\lfloor D^2/2\rfloor$.
The ground state at each $J'/J$
for each bond dimension
is determined by
comparing the energies of candidate states
and taking the lowest-energy state.

\section{Results}
\label{sec:results}

\subsection{Energetics}
\label{subsec:energetics}

\begin{figure}[t]
\centering
\includegraphics[width=0.9\columnwidth]{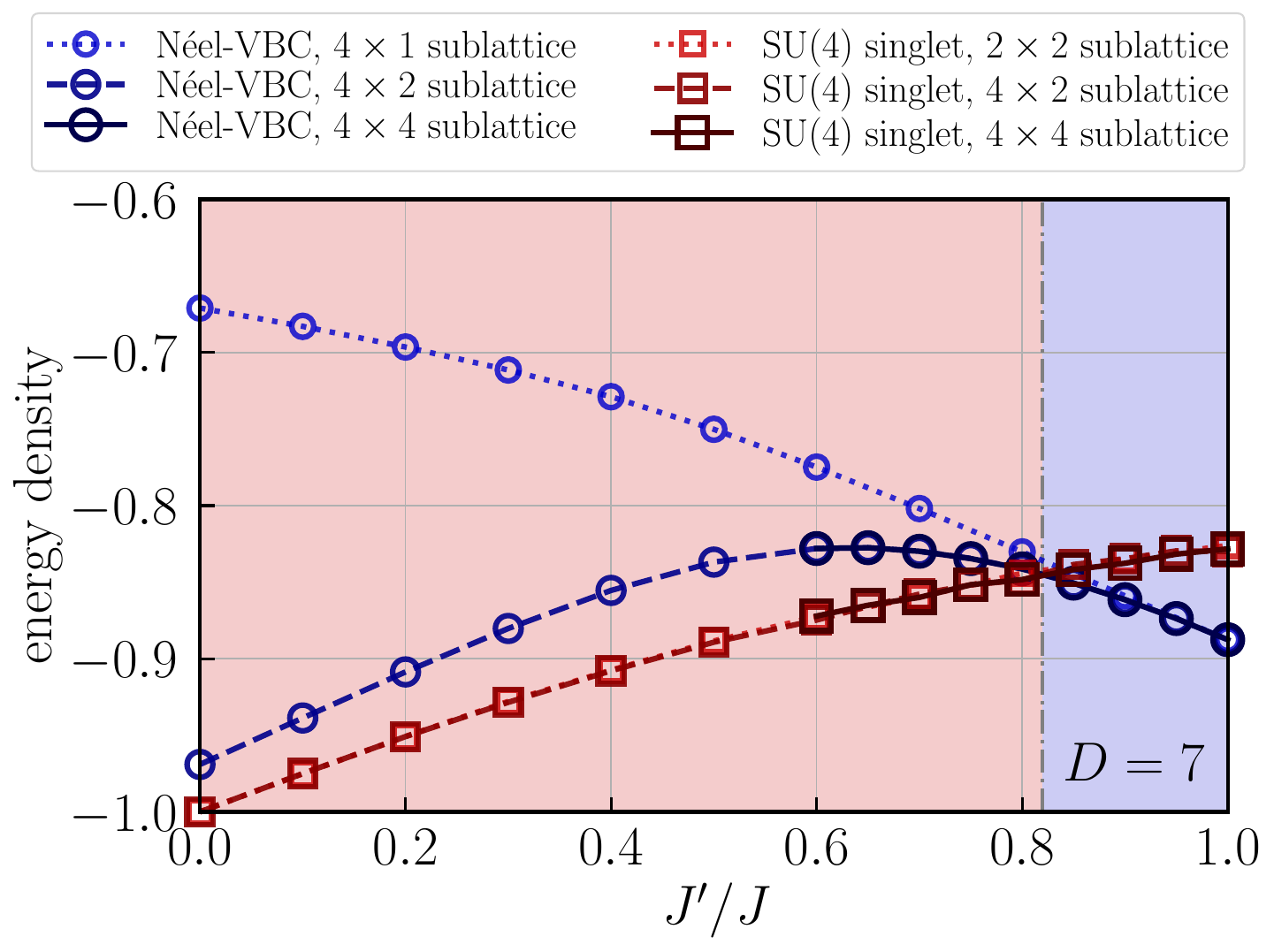}
\caption{%
Sublattice structure dependence of the energy per site
as a function of $J'/J$
for the bond dimension $D=7$.
Squares represent the SU($4$) singlet state
and circles represent the N\'eel-VBC coexisting state.
The dash-dotted line represents the transition point.
}
\label{fig:ene_D7}
\end{figure}

We first investigate the ground-state candidate states
of the model for several sublattice structures
by fixing the bond dimension as $D=7$
(see Fig.~\ref{fig:ene_D7}).

When the SU($4$) singlet state is the initial state for the ground-state optimization process,
the optimized state shows monotonically increasing energy
as $J'/J$ increases.
The energy is nearly the same irrespective of the sublattice structure.
This result indicates that the $2\times 2$ sublattice structure
is suitable for representing the SU($4$) singletlike state,
even for larger $J'/J$.

By contrast, when using the dimerized initial state,
the energy of the optimized state exhibits a sublattice dependence
significantly, in particular, for smaller $J'/J$.
The energy obtained by the $4\times 1$ sublattice structure
is much higher than that obtained by the other sublattice structures,
whereas the energies obtained by the $4\times 2$ and $4\times 4$
sublattice structures are nearly the same.
This observation suggests that
at least the $4\times 2$ sublattice structure
is required for effectively representing the N\'eel-VBC coexistinglike state
for smaller $J'/J$,
while the energy is well converged
by choosing the $4\times 2$ sublattice structure.

\begin{figure}[t]
\centering
\includegraphics[width=0.45\columnwidth]{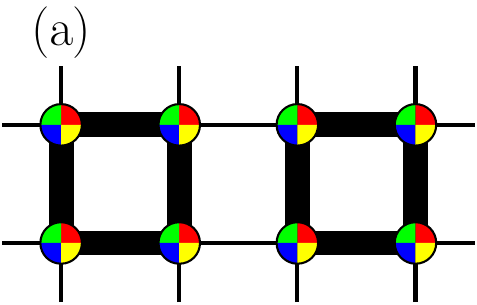}
\hfil
\includegraphics[width=0.45\columnwidth]{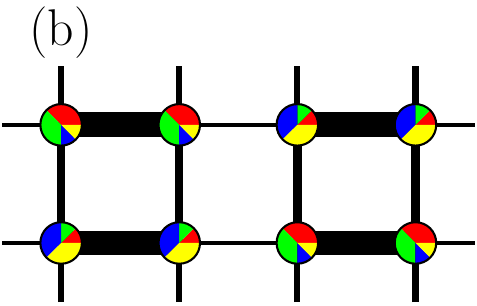}
\caption{%
Schematic pictures of the optimized states.
(a) SU($4$) singlet state for $J'/J\approx 0$.
(b) N\'eel-VBC coexisting state for $J'/J\approx 1$.
Each circle represents a site
and the thickness of the line represents the strength of
the bond correlation $\langle \hat{P}_{ij} \rangle$
between sites $i$ and $j$.
Four colors in each circle represent the four flavors
$\alpha=0,1,2,3$,
and the area of arc represents the ratio of the flavor occupation
$\langle \hat{S}^{\alpha}_{\alpha}(i) \rangle$
at site $i$.
In the absence of
antiferromagnetic order (color order),
$\langle \hat{S}^{\alpha}_{\alpha} \rangle$'s
are equal for all flavors $\alpha=0,1,2,3$,
and they become $1/4$ since
$\sum_{\alpha} \langle \hat{S}^{\alpha}_{\alpha} \rangle=1$.
Color order appears when they deviate from $1/4$.
}
\label{fig:schematic_gs}
\end{figure}

As we will see in more detail later,
for each finite bond dimension,
the state obtained after optimizing the SU($4$) singlet state
still exhibits the SU($4$) singletlike tetramer order
and a small antiferromagnetic order (color order)
even for larger $J'/J$
[see Fig.~\ref{fig:schematic_gs}(a)].
This state nearly keeps the $\mathbb{Z}_4$ lattice rotational symmetry.
On the other hand,
the state obtained after optimizing the dimerized state
for $4\times 2$ or $4\times 4$ sublattice structures
exhibits both dimer and antiferromagnetic order
up to smaller $J'/J$
[see Fig.~\ref{fig:schematic_gs}(b)];
on one dimer bond,
two out of four flavors are dominantly occupied,
whereas on the other dimer bond,
the other two flavors are dominantly occupied.
This dimerized (N\'eel-VBC coexisting)
state spontaneously breaks the lattice
rotational symmetry.

We do not observe other ground-state candidate states
as lowest-energy states in the intermediate parameter region
($0<J'/J<1$) up to the largest $4\times 4$ sublattice structure
at the bond dimension $D=7$.
Therefore, we mainly investigate the competition between the
SU($4$) singlet and dimerized states by increasing bond dimensions.
Because the choice of the $4\times 2$ and $4\times 4$
sublattice structures does not significantly change the optimized states,
hereafter we focus on the result on the $4\times 2$
sublattice structure
to reduce the computational cost.

\begin{figure}[t]
\centering
\includegraphics[width=0.9\columnwidth]{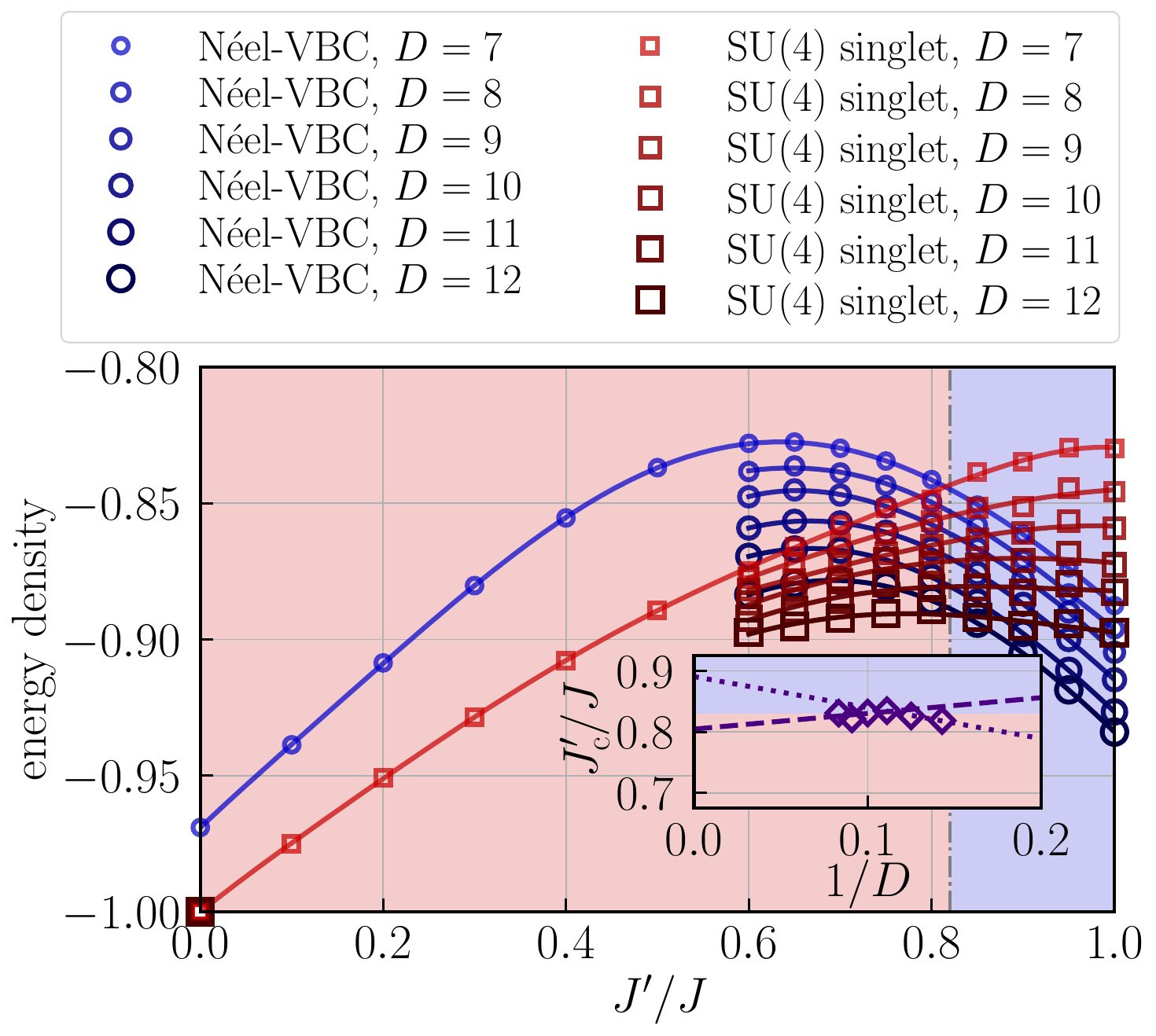}
\caption{%
Bond dimension dependence of the energy per site
as a function of $J'/J$
for the $4\times 2$ sublattice structure.
Squares represent the SU($4$) singlet state
and circles represent the N\'eel-VBC coexisting state.
The energy obtained by a larger bond dimension
is shown by a larger symbol.
The dash-dotted line represents the transition point.
Inset:
The transition point as a function of the bond dimension.
}
\label{fig:ene_4x2}
\end{figure}

We then investigate the energetics
of these two ground-state candidate states
for the $4\times 2$ sublattice structure
by varying the bond dimension $D$.
As shown in Fig.~\ref{fig:ene_4x2},
the energy of the SU($4$) singlet state
gradually decreases as the bond dimension increases.
As the bond dimension increases,
the $J'/J$ dependence of energy becomes weaker
for $J'/J\in [0.6,1]$.
In a similar manner,
the energy of the N\'eel-VBC coexisting state
moderately decreases as the bond dimension increases.
For a fixed bond dimension,
the energy of the N\'eel-VBC coexisting state
signals a peak around $J'/J\approx 0.65$
and monotonically decreases as $J'/J$ increases.
Its curvature does not significantly change
for all bond dimensions $D\ge 7$.

At each bond dimension,
the energy of the SU($4$) singlet state
and that of the N\'eel-VBC coexisting state
cross each other around $J'/J\approx 0.83$.
With increasing bond dimensions,
the transition point is gradually shifted to larger $J'/J$
for $D\in [7,9]$, while it is nearly converged
for $D\ge 10$.
We estimate the transition point
via the linear extrapolation of the crossing points
for smaller $D=7,8,9$ and
those for larger $D=10,11,12$.
From these results,
we conclude that the transition point is around $0.85(5)$
(see the inset of Fig.~\ref{fig:ene_4x2}).

Within the range of bond dimensions, $D\in [7,12]$,
that we have investigated,
the nature of the transition is likely to be of first order.
There are always metastable states of
the SU($4$) singlet state and
the N\'eel-VBC coexisting state near the transition point,
and a level crossing of these two states
is observed for any bond dimensions.
Although we cannot completely
exclude the possibility of a continuous transition
in the infinite bond dimension limit,
investigating the nature of the transition
for larger bond dimensions is
extremely challenging
and we leave it for future study.

\subsection{Physical properties}
\label{subsec:physical_properties}

Having confirmed the phase transition between the SU($4$) singlet
and N\'eel-VBC coexisting states,
we investigate the detailed physical properties
that one would observe in experiments.

\begin{figure}[t]
\centering
\includegraphics[width=0.9\columnwidth]{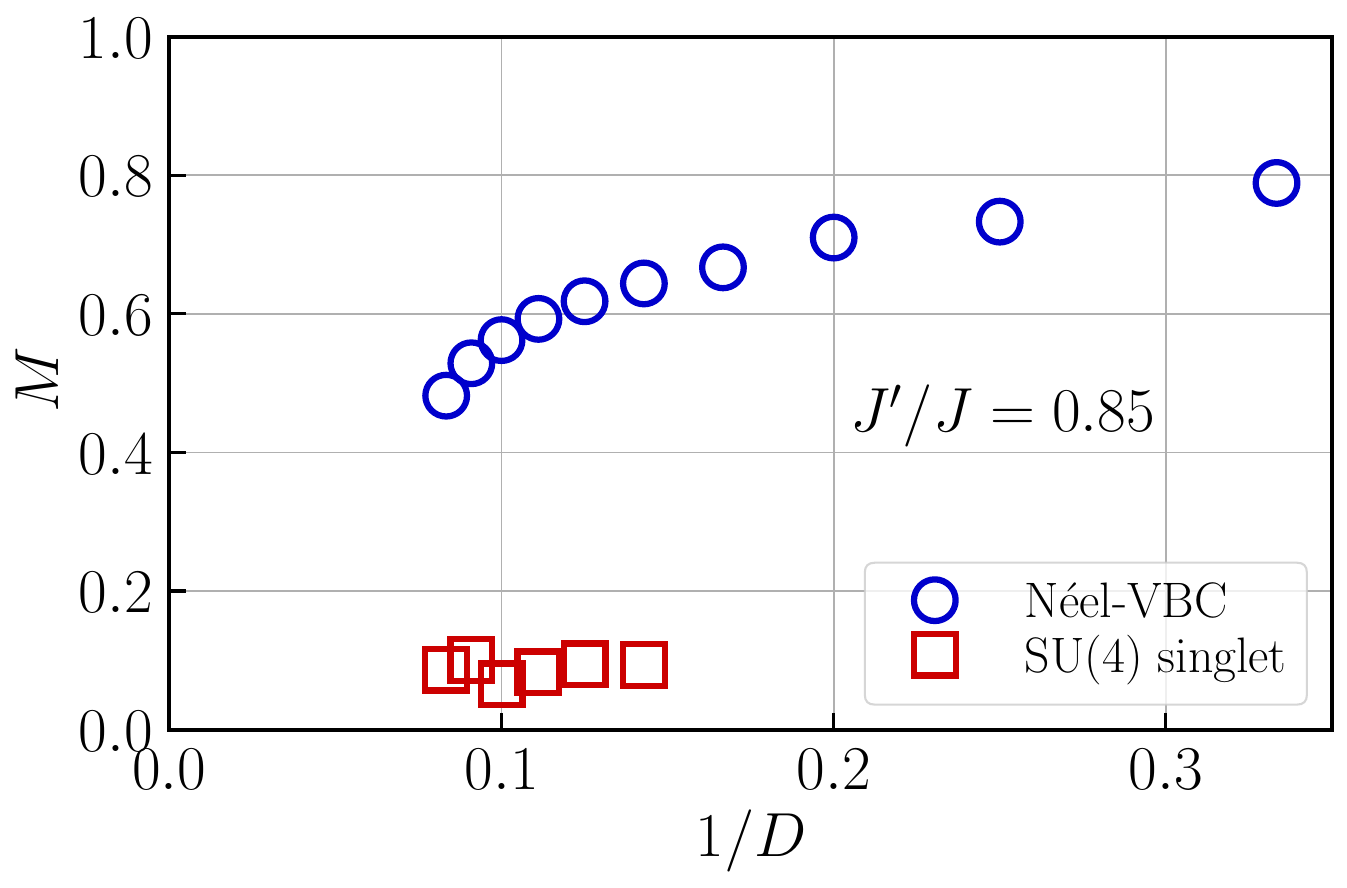}
\caption{%
Bond dimension dependence of the color order parameter
for the SU($4$) singlet (red squares)
and N\'eel-VBC coexisting (blue circles) states
in the $4\times 2$ sublattice structure.
The N\'eel-VBC coexisting state exhibits a sizable color order,
whereas the SU($4$) singlet state shows a small color order.
}
\label{fig:mag}
\end{figure}

We calculate the color order parameter~\cite{corboz2011b},
\begin{align}
 M =
 \frac{1}{N_{\rm sub}} \sum_{i} \sum_{\alpha}
 \left| \langle \hat{S}^{\alpha}_{\alpha}(i) \rangle - \frac{1}{4} \right|,
\end{align}
where $N_{\rm sub}$ is the number of the sublattice sites.
Here the summation is taken over all sublattice sites $i$
and all flavors $\alpha=0,1,2,3$.
In the ideal SU($4$) singlet state at $J'/J=0$,
the color order parameter is zero
because all flavors are equally occupied
($\langle \hat{S}^{\alpha}_{\alpha} \rangle=1/4$).
On the other hand, in the N\'eel-VBC coexisting state at $J'/J=1$,
the color order parameter is nonzero~\cite{corboz2011b}.
Note that in practical iPEPS calculations,
the direction of the SU($4$) symmetry breaking
in the space of four flavors
is explicitly chosen by the initial state
preparation~\cite{corboz2011b}.

Figure~\ref{fig:mag} shows the color order parameter
as a function of bond dimension $D$
for the SU($4$) singlet and N\'eel-VBC coexisting states
near the transition point $J'/J\approx 0.85$.
The N\'eel-VBC coexisting state exhibits a sizable color order
and its value gradually decreases as the bond dimension increases.
On the other hand, the SU($4$) singlet state shows
a much smaller color order than the N\'eel-VBC coexisting state.
Although it is difficult to estimate color order parameters
of both states in the infinite bond dimension limit,
it is likely that the color order parameter
for the SU($4$) singlet state
becomes nearly zero in the infinite bond dimension limit.

\begin{figure}[t]
\centering
\includegraphics[width=0.5\columnwidth]{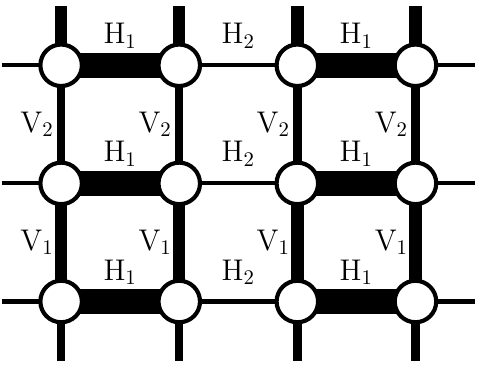}
\caption{%
Name of each bond.
We calculate the bond energies
along the
H$_1$, H$_2$, V$_1$, and V$_2$
bonds
in the $4\times 2$ sublattice structure.
}
\label{fig:def_bond}
\end{figure}

\begin{figure}[t]
\centering
\includegraphics[width=0.9\columnwidth]{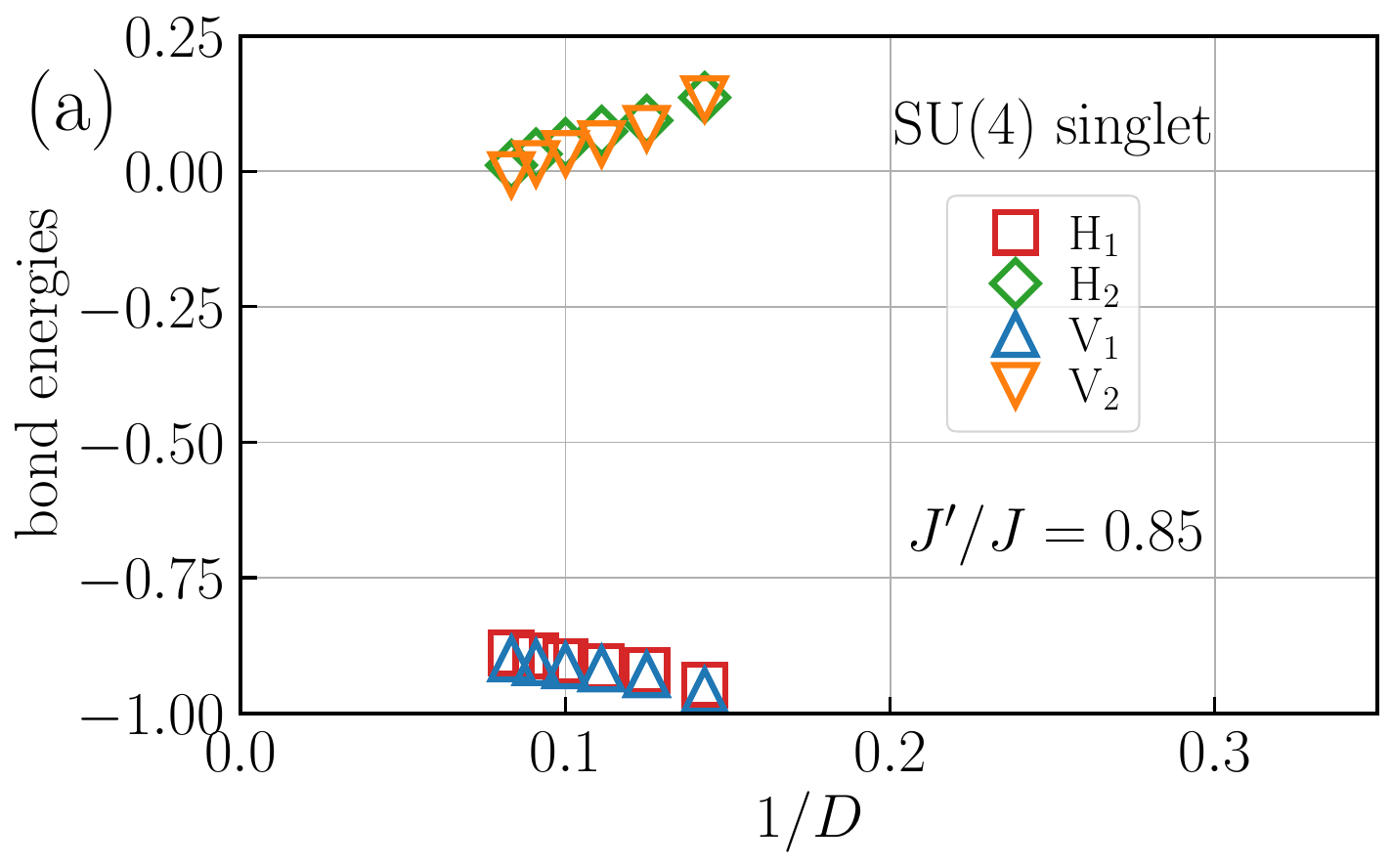}
\\
\includegraphics[width=0.9\columnwidth]{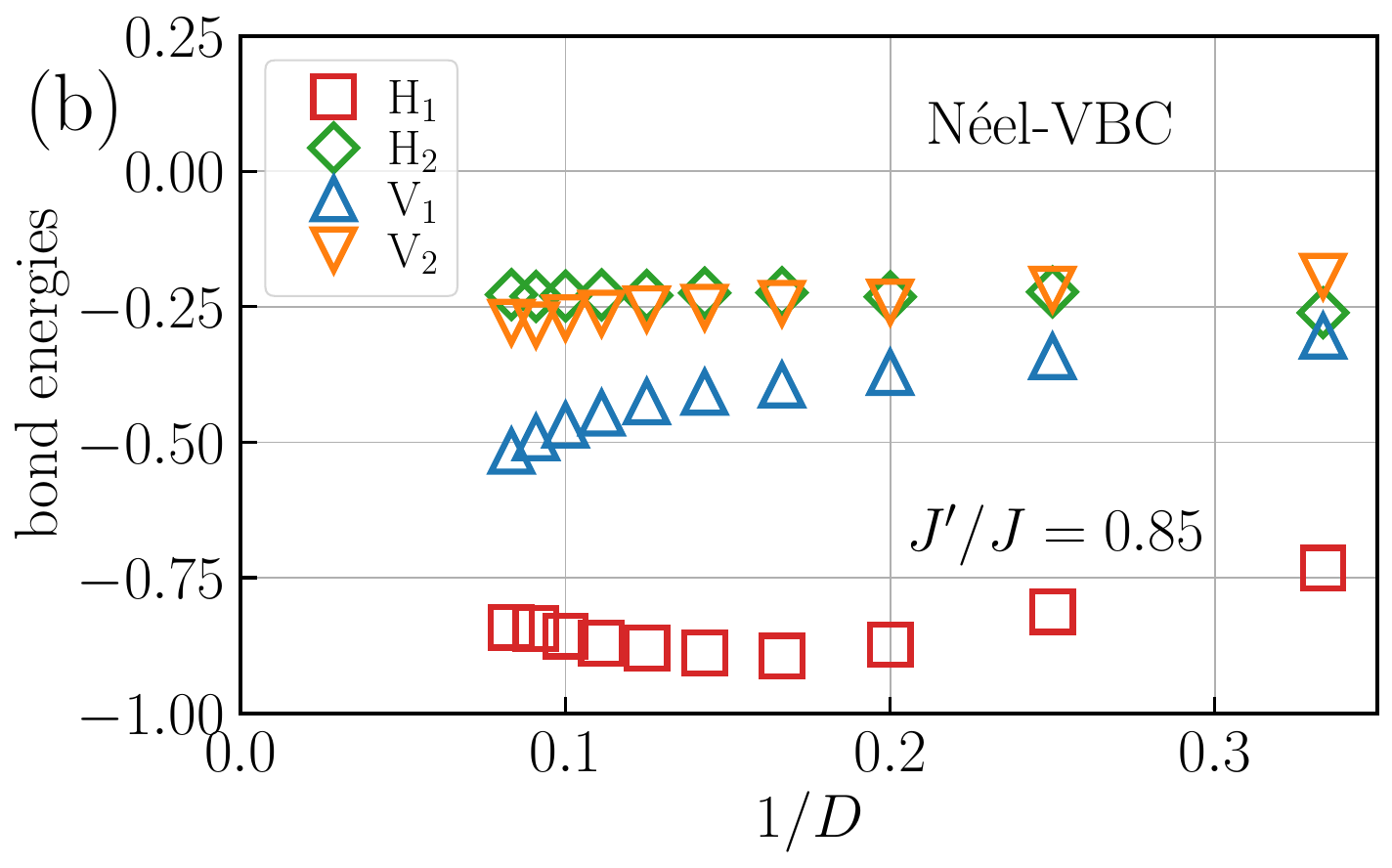}
\caption{%
Bond dimension dependence of each bond energy
for (a) the SU($4$) singlet
and (b) N\'eel-VBC coexisting states
in the $4\times 2$ sublattice structure.
The bond energies are calculated along the
H$_1$ (red squares),
H$_2$ (green diamonds),
V$_1$ (blue up-pointing triangles), and
V$_2$ (orange down-pointing triangles) bonds.
Even near the transition point $J'/J\approx 0.85$,
the SU($4$) singlet state satisfies
$E_{\rm H_1} \approx E_{\rm V_1}$
and
$E_{\rm H_2} \approx E_{\rm V_2}$,
indicating the stable tetramerization.
On the other hand,
the N\'eel-VBC coexisting state exhibits
the stable dimerization
($|E_{\rm H_1}| > |E_{\rm V_1}|,
|E_{\rm V_2}|, |E_{\rm H_2}|$).
}
\label{fig:bond_ene}
\end{figure}

We also examine the bond energies,
\begin{align}
 E_{b} = \frac{1}{N_{b}} \sum_{\langle ij\rangle \in b}
 \langle \hat{P}_{ij} \rangle,
\end{align}
where the symbol $\sum_{\langle ij\rangle \in b}$ denotes the sum over bonds $b$,
and $N_{b}$ is the number of corresponding bonds in the unit cell.
We specifically focus on the bonds
H$_1$, H$_2$, V$_1$, and V$_2$
(see Fig.~\ref{fig:def_bond})
to distinguish the SU($4$) singlet
and N\'eel-VBC coexisting states.

Figure~\ref{fig:bond_ene} shows the bond energies
along the H$_1$, H$_2$, V$_1$, and V$_2$ bonds
for the SU($4$) singlet and N\'eel-VBC coexisting states
near the transition point $J'/J\approx 0.85$.
The SU($4$) singlet state exhibits nearly the same bond energies
along the H$_1$ and V$_1$,
as well as nearly the same bond energies
along the H$_2$ and V$_2$ bonds.
This observation indicates that the SU($4$) singlet state
satisfies the stable tetramerization
even near the transition point.
Considering the fact that the SU($4$) singlet state
shows a nearly zero color order parameter,
the ground state at $J'/J=0$ appears to persist
rather close to the isotropic limit $J'/J=1$.
On the other hand,
the N\'eel-VBC coexisting state exhibits
the stable dimerization
($|E_{\rm H_1}| > |E_{\rm V_1}|,
|E_{\rm V_2}|, |E_{\rm H_2}|$).
Remarkably, the bond energy along the V$_1$ bond
is somewhere in between the largest bond energy along the H$_1$ bond
and the smallest bond energy along the H$_2$ and V$_2$ bonds.
In this sense, the N\'eel-VBC coexisting state
gradually becomes the SU($4$) singletlike state
as $J'/J$ decreases,
although the color order parameter is nonzero
and the lattice rotational symmetry is not fully recovered
even at $J'/J=0$
(not shown).

\begin{figure}[t]
\centering
\includegraphics[width=0.9\columnwidth]{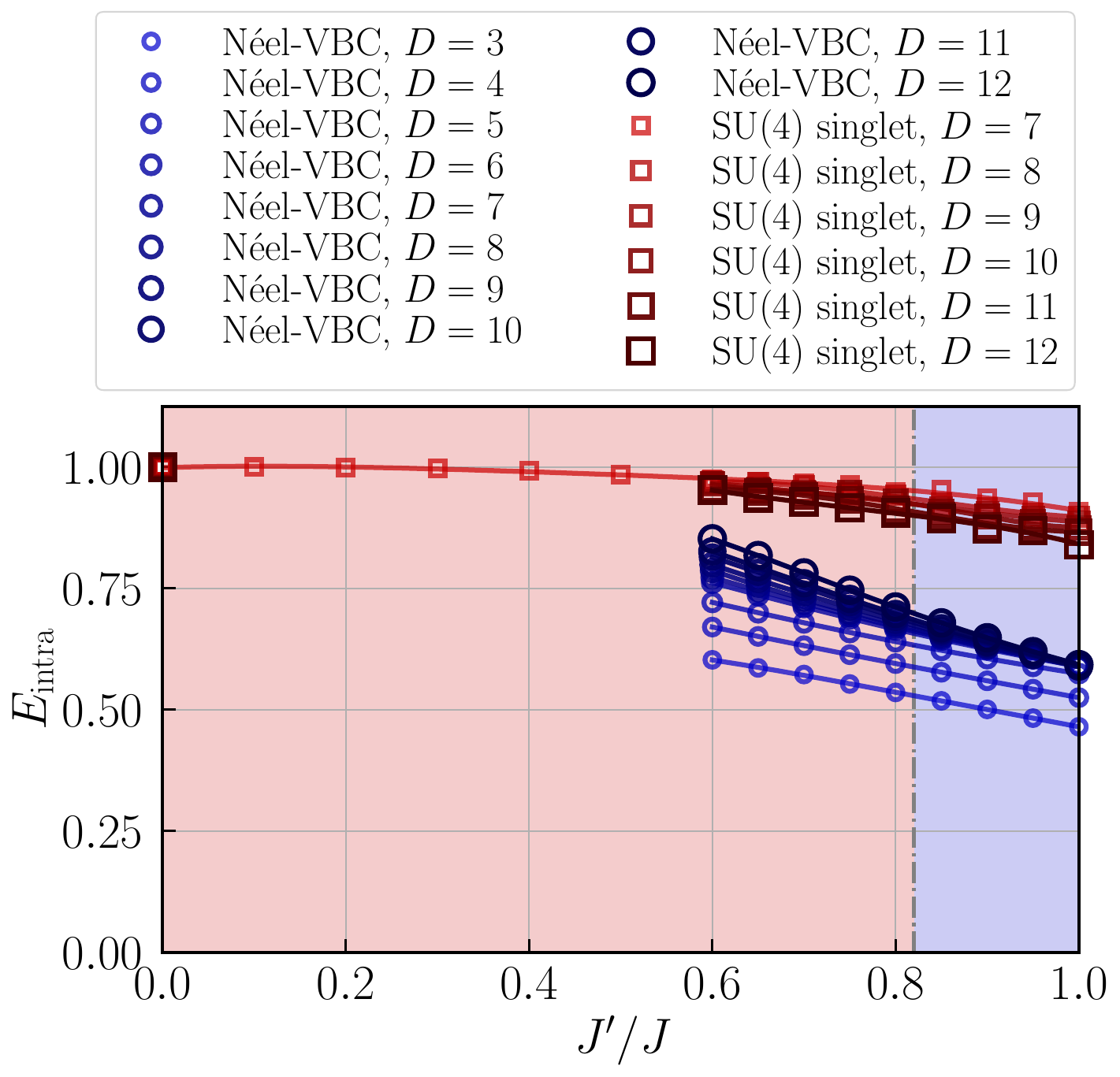}
\caption{%
Anisotropy dependence of
the bond energy on intraplaquette bonds
for the $4\times 2$ sublattice structure.
Squares represent the SU($4$) singlet state
and circles represent the N\'eel-VBC coexisting state.
The data obtained by a larger bond dimension
are shown by a larger symbol.
The dash-dotted line represents the transition point.
}
\label{fig:correlation}
\end{figure}

In SU($N$) systems realized with ultracold atoms in optical lattices,
one multimerizes the system along specific directions
and observes an excess of singlets compared with triplets.
By optically inducing a singlet-triplet oscillation~\cite{trozky2010},
one can measure antiferromagnetic correlations~\cite{taie2012,ozawa2018,
taie2022}.
To directly compare with such quantities accessible in experiments,
we also calculate the bond energy only on intraplaquette bonds,
which is defined as
\begin{align}
 E_{\rm intra}
 = -\frac{1}{N_{\rm intrabond}} \sum_{\langle ij \rangle_{\rm intra}}
 \langle \hat{P}_{ij} \rangle
 = - \frac{E_{\rm H_1} + E_{\rm V_1}}{2}.
\end{align}
Here, $N_{\rm intrabond}$ is the number of the intraplaquette bonds
in the unit cell,
and $N_{\rm intrabond}=8$ is
for the $4\times 2$ sublattice structure.
It takes the value $E_{\rm intra}=1$
when the ground state is the SU($4$) singlet state
and, in general, measures the intensity of singlet correlations
along the intraplaquette bonds.

We illustrate the anisotropy dependence of the bond energy
on intraplaquette bonds as shown in Fig.~\ref{fig:correlation}.
For a smaller $J'/J\lesssim 0.83$,
the ground state is the SU($4$) singlet state
and its bond energy is nearly saturated.
This observation indicates that
the stable tetramerization persists
even for large $J'/J$ close to the isotropic limit.
For $J'/J\gtrsim 0.83$,
the ground state becomes the N\'eel-VBC coexisting state
and its bond energy is smaller than that of the SU($4$) singlet state.
As $J'/J$ increases,
the dimerization becomes stronger
and the bond energy along the vertical bonds V$_1$
and that along the horizontal bond H$_1$
deviate from each other.
The bond energy on intraplaquette bonds
jumps around $J'/J\approx 0.83$,
which might be detected in future experiments
of ultracold atoms in optical lattices by
utilizing singlet-triplet oscillations.

\section{Summary and outlook}
\label{sec:summary}

In conclusion, we investigated the ground state of the SU($4$)
Heisenberg model on a plaquette lattice
using the two-dimensional tensor-network method based on iPEPS.
We analyzed two competing ground-state candidate states:
the SU($4$) singlet state and the N\'eel-VBC coexisting state.
We showed that the former is the ground state in the strongly anisotropic limit,
whereas the latter is the ground state in the nearly isotropic limit.
By examining the intermediate parameter region,
we identified a first-order phase transition between these phases.
The transition point was estimated to be around
$J'/J\approx 0.85(5)$, where $J$ and $J'$ are the interaction strengths
of the intra- and interplaquette bonds, respectively.

Compared to the SU($2$) model on a plaquette lattice,
where the transition between
the $s$-wave resonating-valence-bond state~\cite{belen2008,nascimbene2012}
and
the N\'eel state
takes place around $0.5485(2)$~\cite{wenzel2009},
the SU($4$) singlet ground state persists in the nearly isotropic limit.
Indeed, exact diagonalization calculations for $4\times 4$ sites
give the SU($4$) singletlike (plaquette) ground state
on the isotropic square lattice~\cite{vandenbossche2000},
and such a state seems to be a low-energy state (but not the ground state)
in the thermodynamic limit.
This observation suggests that the spontaneous multimerization
is more easily realized in SU($N$) systems with very large $N$.
In this sense, investigating the ground state of the SU($N$) model on an isotropic
square lattice for larger $N$ will be an interesting future study.

In ultracold-atom experiments,
one can measure antiferromagnetic correlations
by optically inducing a singlet-triplet oscillation~\cite{trozky2010}.
We calculated the anisotropy dependence of such correlations
and found that it is nearly saturated even for large $J'/J\lesssim
0.85(5)$.
It eventually jumps around the transition point
and takes a relatively large value
(but smaller than the saturated value)
near the isotropic limit.
These results will be useful for future experiments
with ultracold atoms in optical lattices.

The effects of magnetic field and other perturbations
will enrich the phase diagram of the model~\cite{miyazaki2021,miyazaki2022}.
It is also interesting to investigate the ground state
of the SU($4$) model in such situations
using the tensor-network method,
which we leave for future study.

\begin{acknowledgments}
The authors acknowledge fruitful discussions with
Kentaro Honda,
Daichi Kagamihara,
Mathias Mikkelsen,
Yuki Miyazaki,
Shintaro Taie,
Yoshiro Takahashi,
Yosuke Takasu,
Takuto Tsuno,
and
Daisuke Yamamoto.
This work was supported by
JSPS KAKENHI (Grant No.\ JP24H00973),
MEXT KAKENHI, Grant-in-Aid for Transformative Research Area
(Grants No.\ JP22H05111 and No.\ JP22H05114),
JST FOREST (Grant No.\ JPMJFR202T),
and
MEXT Q-LEAP (Grant No.\ JPMXS0118069021).
The numerical computations were performed on computers at
the Yukawa Institute Computer Facility,
Kyoto University
and on computers at
the Supercomputer Center, the Institute for Solid State Physics,
the University of Tokyo.
\end{acknowledgments}

\appendix

\section{Initial state preparation}
\label{sec:initial_state}

In this section, we describe the initial state preparation
for the iPEPS calculation.
We specifically focus on the SU($4$) singlet initial state
and the dimerized initial state
for the SU($4$) Heisenberg model on a plaquette lattice.

\subsection{Tensor-network representation of the SU($N$) singlet state}
\label{subsec:sun_init}

\begin{table}[!ht]
\caption{Power set of a set $S=\{0, 1, 2, \dots, N-1\}$ for $N=3$
and the corresponding index for each subset.
The symbol \# denotes the number of elements in each set.
The index of the left or right virtual bond takes the value of ${\rm ind}({\rm set})$.}
\label{tab:app:set_index}
\begin{tabular}{ccc}
\hline
 set & \#set & ${\rm ind}({\rm set})$ \\
\hline
\hline
 $\{\}$ & 0 & 0 \\
\hline
 $\{0\}$ & 1 & 0 \\
 $\{1\}$ & 1 & 1 \\
 $\{2\}$ & 1 & 2 \\
\hline
 $\{0,1\}$ & 2 & 0 \\
 $\{0,2\}$ & 2 & 1 \\
 $\{1,2\}$ & 2 & 2 \\
\hline
 $\{0,1,2\}$ & 3 & 0 \\
\hline
\end{tabular}
\end{table}

\begin{figure}[!ht]
\includegraphics[width=.72\columnwidth]{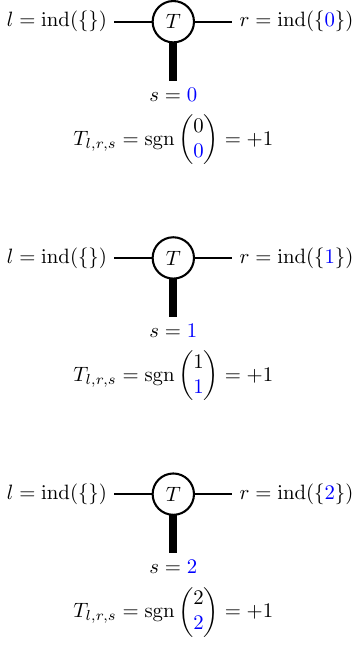}
\caption{Nonzero MPS elements at site $0$ for the ${\rm SU}(N=3)$ singlet state.
The index of the left or right virtual bond takes the value of ${\rm ind}({\rm set})$,
which is defined in Table~\ref{tab:app:set_index}.
The value of the tensor $T_{l,r,s}$ is given by the sign
of the permutation corresponding to a sequence
$(s_0, s_1, \dots, s_{\#{\rm set}_l-1}, s)$,
whose order matters.
The sequence is generated from the set at the left bond
(${\rm set}_l = \{s_0, s_1, \dots, s_{\#{\rm set}_l-1}\}$)
and the color of the physical bond
($s=0$, $1$, or $2$).
At site $0$,
the set at the left bond is the empty set
and that at the right bond is a size-$1$ subset.
}
\label{fig:app:mps_site_0}
\end{figure}

\begin{figure}[!ht]
\includegraphics[width=.72\columnwidth]{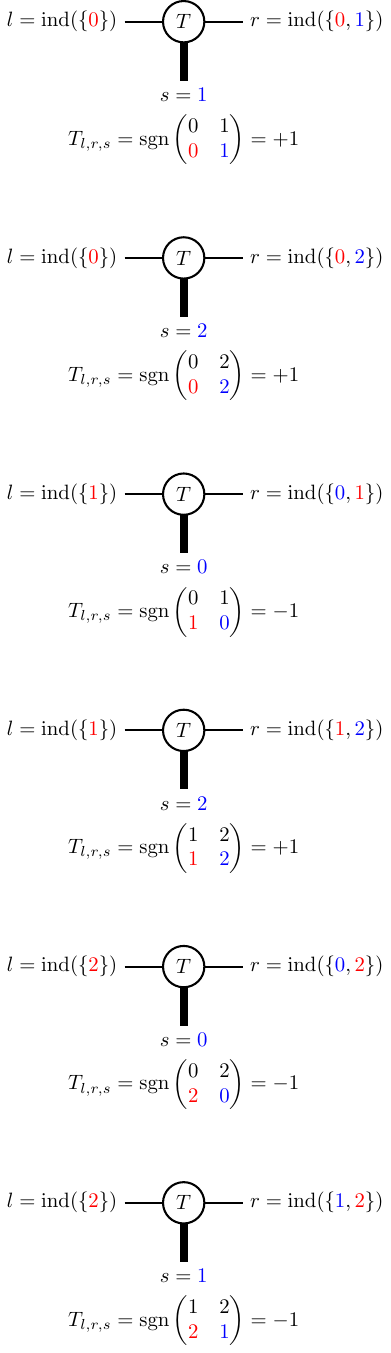}
\caption{Nonzero MPS elements at site $1$ for the ${\rm SU}(N=3)$ singlet state.
The notation is the same as that in Fig.~\ref{fig:app:mps_site_0}.}
\label{fig:app:mps_site_1}
\end{figure}

\begin{figure}[!ht]
\includegraphics[width=.72\columnwidth]{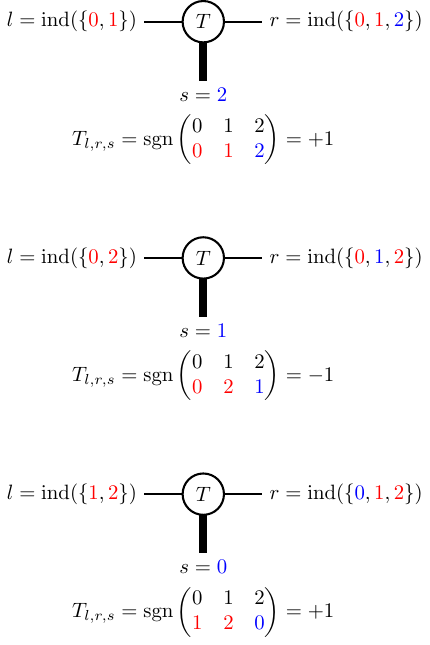}
\caption{Nonzero MPS elements at site $2$ for the ${\rm SU}(N=3)$ singlet state.
The notation is the same as that in Fig.~\ref{fig:app:mps_site_0}.}
\label{fig:app:mps_site_2}
\end{figure}

\begin{table}[!ht]
\caption{Nonzero tensor elements of the iPEPS representation
of the SU($4$) singlet state.
The sublattice sites A, B, C, and D of the $2\times 2$ sublattice structure
defined in Fig.~\ref{fig:sublattice}(a) are represented by
the site index $(x,y)=(0,0)$, $(1,0)$, $(0,1)$, and $(1,1)$, respectively.
The index of the left, top, right, and bottom virtual bonds
takes the value $0,1,\dots,6(=D-1)$ with $D=7$.
The index of the physical bond (color)
$0,1,2,3$ corresponds to the four flavors
of the SU($4$) model.
The value of the tensor $T_{\rm left,top,right,bottom,color}$ is given
by the appropriate sign.
This iPEPS representation on a plaquette
corresponds to the MPS representation on a four-site chain
(connecting sites in order of A, B, D, and C)
with the bond dimension $D_{\rm MPS}=6$
(corresponding to the index of virtual bonds $1,2,\dots,6$).
}
\label{tab:ipeps_elements_singlet}
\centering%
\begin{tabular}{ccccccc}
\hline
site & left & top & right & bottom & color & value \\
\hline
\hline
(0,0) & 0 & 1 & 1 & 0 & 0 & $+1$ \\
(0,0) & 0 & 1 & 2 & 0 & 1 & $+1$ \\
(0,0) & 0 & 1 & 3 & 0 & 2 & $+1$ \\
(0,0) & 0 & 1 & 4 & 0 & 3 & $+1$ \\
(1,0) & 1 & 1 & 0 & 0 & 1 & $+1$ \\
(1,0) & 1 & 2 & 0 & 0 & 2 & $+1$ \\
(1,0) & 1 & 3 & 0 & 0 & 3 & $+1$ \\
(1,0) & 2 & 1 & 0 & 0 & 0 & $-1$ \\
(1,0) & 2 & 4 & 0 & 0 & 2 & $+1$ \\
(1,0) & 2 & 5 & 0 & 0 & 3 & $+1$ \\
(1,0) & 3 & 2 & 0 & 0 & 0 & $-1$ \\
(1,0) & 3 & 4 & 0 & 0 & 1 & $-1$ \\
(1,0) & 3 & 6 & 0 & 0 & 3 & $+1$ \\
(1,0) & 4 & 3 & 0 & 0 & 0 & $-1$ \\
(1,0) & 4 & 5 & 0 & 0 & 1 & $-1$ \\
(1,0) & 4 & 6 & 0 & 0 & 2 & $-1$ \\
(1,1) & 1 & 0 & 0 & 1 & 2 & $+1$ \\
(1,1) & 2 & 0 & 0 & 1 & 3 & $+1$ \\
(1,1) & 1 & 0 & 0 & 2 & 1 & $-1$ \\
(1,1) & 3 & 0 & 0 & 2 & 3 & $+1$ \\
(1,1) & 2 & 0 & 0 & 3 & 1 & $-1$ \\
(1,1) & 3 & 0 & 0 & 3 & 2 & $-1$ \\
(1,1) & 1 & 0 & 0 & 4 & 0 & $+1$ \\
(1,1) & 4 & 0 & 0 & 4 & 3 & $+1$ \\
(1,1) & 2 & 0 & 0 & 5 & 0 & $+1$ \\
(1,1) & 4 & 0 & 0 & 5 & 2 & $-1$ \\
(1,1) & 3 & 0 & 0 & 6 & 0 & $+1$ \\
(1,1) & 4 & 0 & 0 & 6 & 1 & $+1$ \\
(0,1) & 0 & 0 & 1 & 1 & 3 & $+1$ \\
(0,1) & 0 & 0 & 2 & 1 & 2 & $-1$ \\
(0,1) & 0 & 0 & 3 & 1 & 1 & $+1$ \\
(0,1) & 0 & 0 & 4 & 1 & 0 & $-1$ \\
\hline
\end{tabular}
\end{table}

\begin{table}[!ht]
\caption{Nonzero tensor elements of the iPEPS representation
of the dimerized state.
The sublattice sites A, B, C, and D of the $4\times 1$ sublattice structure
defined in Fig.~\ref{fig:sublattice}(b) are represented by
the site index $(x,y)=(0,0)$, $(1,0)$, $(2,0)$, and $(3,0)$, respectively.
The index of the left, top, right, and bottom virtual bonds
takes the value $0,1,2(=D-1)$ with $D=3$.
The index of the physical bond (color)
$0,1,2,3$ corresponds to the four flavors
of the SU($4$) model.
The value of the tensor $T_{\rm left,top,right,bottom,color}$ is given
by the appropriate sign.
}
\label{tab:ipeps_elements_dimer}
\centering%
\begin{tabular}{ccccccc}
\hline
site & left & top & right & bottom & color & value \\
\hline
\hline
(0,0) & 0 & 0 & 1 & 0 & 0 & $+1$ \\
(0,0) & 0 & 0 & 2 & 0 & 1 & $+1$ \\
(1,0) & 1 & 0 & 0 & 0 & 1 & $+1$ \\
(1,0) & 2 & 0 & 0 & 0 & 0 & $-1$ \\
(2,0) & 0 & 0 & 1 & 0 & 2 & $+1$ \\
(2,0) & 0 & 0 & 2 & 0 & 3 & $+1$ \\
(3,0) & 1 & 0 & 0 & 0 & 3 & $+1$ \\
(3,0) & 2 & 0 & 0 & 0 & 2 & $-1$ \\
\hline
\end{tabular}
\end{table}

Here we show that the ${\rm SU}(N)$ singlet state
on an $N$-site chain can be described by the matrix product state (MPS)
with the maximum virtual bond dimension
$D_{\rm MPS}=\binom{N}{\lfloor N/2\rfloor}$.
For simplicity, let us focus on the ${\rm SU}(N=3)$ case,
where the required maximum bond dimension is $D_{\rm MPS}=3$.
The singlet state is given by
\begin{align}
 |\psi\rangle
 &\propto
  |012\rangle
 -|021\rangle
 -|102\rangle
 +|120\rangle
 +|201\rangle
 -|210\rangle
\\
 &=
 \sum_{i,j,k \in \{0,1,2\}}
 \epsilon_{i,j,k} |ijk\rangle,
\end{align}
where $\epsilon_{i,j,k}$ denotes the Levi-Civita symbol.
We will construct the MPS at each site ($0$, $1$, and $2$) in a way such that
the trace of the product of the MPSs gives $\epsilon_{i,j,k}$.

To this end,
we first prepare the power set of a set $S=\{0, 1, 2, \dots, N-1\}$.
For each subset containing $n(\in \{0, 1, 2, \dots, N-1, N\})$ elements,
we define the function ${\rm ind}(\cdot)$
as shown in Table~\ref{tab:app:set_index}.
For example, to generate the subset for $\rm \#set=2$,
we select the smallest number $0$ as the first element
and then select the second smallest number $1$ as the second element,
which gives the subset $\{0,1\}$.
To avoid double counting,
we enumerate the subsets in ascending order
so that the right element is always larger than the left element.
We assign the index $0$ to this subset $\{0,1\}$.
Next, we fix the first element $0$ and select the next smallest number $2$,
which gives the subset $\{0,2\}$. The index of this subset is $1$.
We repeat this procedure by fixing the first element $1$
and increasing the remaining elements
until the last element $N-1$ is selected.
We then select the second smallest number $1$ as the first element
and select the next smallest number $2$ as the second element,
which gives the subset $\{1,2\}$. The index of this subset is $2$.
Again, we fix the first element and increase the remaining elements
until the last element $N-1$ is selected.
We repeat this procedure until we obtain
the last subset $\{N-n,\dots,N-3,N-2,N-1\}$
for ${\rm \#set} = n$ in general.
(For $N=3$ and $n=2$, the subset $\{1,2\}$
is the last subset.)
Then, the maximum index is given by $\binom{N}{n}-1$.

The index ${\rm ind}({\rm set})$
will be used to label the left and right virtual bonds.
On the other hand,
the physical bond takes the flavor $s=0$, $1$, or $2$,
and so does its index.
This choice corresponds to providing $s$ in ascending order
from the elements of the set $\{0,1,2\}$ satisfying ${\rm \#set}=1$.
As we will see below,
the element of the tensor $T_{l,r,s}$
is chosen as the sign of the permutation corresponding to
a certain sequence, whose order matters,
that is generated from the set at the left bond
and the color at the physical bond.

Next, we construct the MPS at site $0$ (see Fig.~\ref{fig:app:mps_site_0}).
For the left bond, we choose an empty set ${\rm set}_l=\{\}$.
The maximum dimension of the left bond is $1$ at site $0$,
and the index of the left bond can only take the value $l=0$.
For the physical bond, the color can take the value
$s=0$, $1$, or $2(=N-1)$.
For the right bond, we choose a set
depending on the color at the physical bond, i.e., ${\rm set}_r=\{s\}$.
The index of the right bond takes the value $r={\rm ind}({\rm set}_r)=s$
and the maximum dimension of the right bond is $3$ at site $0$.
The value of the tensor element $T_{l,r,s}$ at site $0$ will be
equivalent to $T_{0,s,s}$. It is chosen as $+1$ for all $s$
because the permutation corresponding to the sequence $(s)$
is the identity.

Then, we construct the MPS at site $1$ (see Fig.~\ref{fig:app:mps_site_1}).
For the left bond, we choose a set
corresponding to that for the right bond at site $0$.
It is given as ${\rm set}_l=\{s_0\}$, with $s_0=0$, $1$, or $2$
being the color of the physical bond at site $0$.
The maximum dimension of the left bond is $3$ at site $1$,
and the index of the left bond takes the value $l=0$, $1$, or $2$.
For the right bond, we prepare a set
${\rm set}_r=\{s_0,s\}$,
which is constructed from the element of ${\rm set}_l=\{s_0\}$
and the color at the physical bond $s$ at site $1$.
Since the index of the right bond takes the value $r=0$, $1$, or $2$,
the maximum dimension of the right bond is $3$ at site $1$.
The tensor element $T_{l,r,s}$ at site $1$ is nonzero only when $\# {\rm set}_r=2$.
It will be
obtained by the sign of the permutation corresponding to a sequence
$(s_0,s)$, whose order matters;
it is positive (negative) if the parity is even (odd).
For example, when we have a sequence $(0,1)$,
we generate the permutation
$\sigma =
\begin{pmatrix}
 0 & 1 \\
 0 & 1 \\
\end{pmatrix}
$ and assign its sign ${\rm sgn}(\sigma)=+1$ to $T_{l,r,s}$.
On the other hand,
when we have a sequence $(1,0)$,
we generate the permutation
$\sigma =
\begin{pmatrix}
 0 & 1 \\
 1 & 0 \\
\end{pmatrix}
$ and assign its sign ${\rm sgn}(\sigma)=-1$ to $T_{l,r,s}$.

Finally, we construct the MPS at site $2$ (see Fig.~\ref{fig:app:mps_site_2}).
For the left bond, we choose a set
corresponding to that for the right bond at site $1$.
It is given as ${\rm set}_l=\{s_0,s_1\}=\{0,1\}$, $\{0,2\}$, or $\{1,2\}$.
The maximum dimension of the left bond is $3$ at site $2$,
and the index of the left bond takes the value $l=0$, $1$, or $2$.
For the right bond, we prepare a set
${\rm set}_r=\{s_0,s_1,s\}$,
which is constructed from the element of ${\rm set}_l=\{s_0,s_1\}$
and the color at the physical bond $s$ at site $2$.
This actually becomes
${\rm set}_r\equiv\{0,1,2\}$,
and then the index of the right bond only takes the value
$r = {\rm ind}(\{0,1,2\}) = 0$.
The corresponding maximum dimension of the right bond is $1$ at site $2$.
This dimension $1$ is consistent with that of the left bond at site $0$,
which allows us to safely calculate the trace of product of three tensors.
The tensor element $T_{l,r,s}$ at site $2$ is nonzero only when $\# {\rm set}_r=3$,
and it will again be
obtained by the sign of the permutation corresponding to a sequence
$(s_0,s_1,s)$.

From these three tensors $T^{(i)}_{l,r,s}$ at sites $i=0$, $1$, and $2$,
we evaluate the trace of product of them. Easy calculations yield
\begin{align}
 \underset{i_0,i_1,i_2}{\rm Tr}\,
 T^{(0)}_{i_0,i_1,s_0}
 T^{(1)}_{i_1,i_2,s_1}
 T^{(2)}_{i_2,i_0,s_2}
 = \epsilon_{s_0,s_1,s_2}.
\end{align}
Here
the virtual bonds take the value $i_0=0$, $i_1,i_2\in\{0,1,2\}$,
and the physical bond takes the value $s_0,s_1,s_2\in\{0,1,2\}$.

One can generalize this construction to the general SU($N$) case.
In the iPEPS representation of the SU($N$) singlet state,
the required virtual bond dimension is
increased by $1$ and given by
$D=D_{\rm MPS}+1=\binom{N}{\lfloor N/2\rfloor}+1$.
For $N=4$, the required bond dimension is $D=7$
and the corresponding tensor elements of the iPEPS
are summarized in Table~\ref{tab:ipeps_elements_singlet}.

\subsection{Tensor-network representation of the dimerized state}
\label{subsec:dimer_init}

We prepare the dimerized initial state
by placing the singlet pairs along the horizontal bonds
connecting the sublattice sites A and B
and those connecting the sublattice sites C and D
in the $4\times 1$ sublattice structure
[see Fig.~\ref{fig:sublattice}(b)
and Fig.~\ref{fig:schematic_gs}(b)].
For one singlet pair,
we only use the physical bonds $0$ and $1$,
corresponding to two out of four flavors of the SU($4$) model.
For the other singlet pair,
we only use the physical bonds $2$ and $3$,
corresponding to the remaining two flavors.
Resulting tensor elements of the iPEPS
are summarized in Table~\ref{tab:ipeps_elements_dimer}.
The required virtual bond dimension of this initial state
in the iPEPS representation is $D=3$~\cite{verstraete2006,schuch2012}.

\bibliographystyle{apsrev4-2}

\onecolumngrid

\end{document}